\documentclass[10pt]{iopart}
\usepackage{graphicx}
\usepackage{hyperref}
\usepackage{cite}
\usepackage{xcolor}
\expandafter\let\csname equation*\endcsname\relax
\expandafter\let\csname endequation*\endcsname\relax
\usepackage{amsmath}
\usepackage{amsfonts}
\usepackage{amssymb}
\usepackage{amsthm}
\begin{document}
\title{Channeling of fluorescence photons from quantum dots into guided modes of an optical nanofiber tip}
\author{Resmi M, Elaganuru Bashaiah, and Ramachandrarao Yalla}
\address{School of Physics, University of Hyderabad, Hyderabad, Telangana, India-500046}
\ead{rrysp@uohyd.ac.in}
\vspace{10pt}
\begin{indented}
\item[]January 2024
\end{indented}
\begin{abstract}
We demonstrate the channeling of fluorescence photons from quantum dots (QDs) into guided modes of an optical nanofiber tip (ONFT). We deposit QDs on the ONFT using micro/nano fluidic technology. We measure the photon-counting and emission spectrum of fluorescence photons that are channeled into guided modes of the ONFT. The measured emission spectrum confirms the deposition of QDs on the ONFT. We perform numerical simulations to determine channeling efficiency ($\eta$) for the ONFT and a single dipole source (SDS) system. For the radially oriented SDS at the center of the facet of the ONFT, we found the maximum $\eta$-value of 44\% at the fiber size parameter of 7.16, corresponding to the ONFT radius of 0.71 $\mu m$ for the emission wavelength at 620 $nm$. Additionally, we investigate the SDS position dependence in transverse directions on the facet of the ONFT in view of keeping experimental ambiguities. The present fiber inline platform may open new avenues in quantum technologies.

\end{abstract}
\vspace{2pc}
\noindent{\it Keywords}: Optical nanofiber tip, Single photons, Quantum dots.
\ioptwocol
\section{Introduction}
The ultimate control of light-matter interaction at the single photon level and the implementation of such a system are challenging tasks in contemporary quantum physics, with potential applications in the field of quantum technologies\cite{kimble2008quantum,slussarenko2019photonic}. Single photons are ideal information carriers for quantum information science including quantum networks. Moreover, with the present technology, the information can be readily encoded into the photon degrees of freedom for exchange between distant quantum processing nodes, leading to quantum entanglement, another key ingredient of quantum technology. Many protocols have been proposed and implemented to achieve efficient manipulation of single photons using various techniques. Examples would include high finesse optical cavities in free-space\cite{hood2000atom}, photonic crystal cavities\cite{schroder2011ultrabright}, micropillar cavities\cite{solomon2001single}, diamond nanobeams\cite{hausmann2013coupling}, and silicon nitride alligator waveguides\cite{hung2013trapped}. 

However, the crucial challenge lies in efficiently directing single photons emitted by a single photon source into single-mode fibers (SMF). To address this challenge, recently optical nanofibers (ONF) \cite{PhysRevLett.109.063602} and optical nanofiber tips (ONFT) \cite{chonan2014efficient,resmi2023efficient} are proven to be promising candidates. ONFs are tapered optical fibers with a sub-wavelength diameter at the waist region. ONFTs are tapered optical fibers with a sub-wavelength diameter at one edge. A substantial difference in effective refractive index between the core (silica) and clad (air) of the ONF/ONFT leads to properties such as strong optical confinement, large evanescent field, strong field enhancement, etc. \cite{tong2004single,tong2011subwavelength}. Due to the large evanescent field, ONFs are indispensable components in the fields of quantum optics, quantum information science \cite{bures1999power}, and sensing \cite{taha2021comprehensive,afshar2007enhancement}. Furthermore, the intensity distribution is modified due to the silica core and vacuum clad \cite{le2004field}. Channeling of the dipole emission into a step-index fiber with respect to different orientations and positions has been modeled by Matthew R. Henderson {\it et al.} \cite{henderson2011dipole}. They demonstrated that the significant refractive index contrast between the core and the clad of the SMF is highly effective for channeling single photons. 

Various groups worldwide have made notable progress in enhancing channeling efficiency ($\eta$). Efficiently channeling single photons emitted by a single photon source becomes achievable when the source is positioned in the vicinity of the ONF, as supported by theoretical studies \cite{klimov2004spontaneous}. The decay rate into each mode of a multi-mode waveguide when a single photon emitter is placed close to its surface and changes in the $\eta$-value with respect to the guide's radius and dipole orientations have been studied \cite{verhart2014single}. Solid state single photon emitters \cite{aharonovich2016solid} such as quantum dots (QD), nitrogen-vacancy (NV) centers, and silicon-vacancy (SiV) centers are being used for experiments. In some experiments ions, dye molecules, and defect centers in diamond crystals are used as single photon emitters \cite{mckeever2004deterministic,keller2004continuous,lounis2000single,michler2000quantum,santori2002indistinguishable,kurtsiefer2000stable}. Also, single photons are generated in a controlled way from a strongly coupled atom-cavity system \cite{kuhn1999controlled} and a single trapped two-level atom \cite{darquie2005controlled}.

A cesium atom in the vicinity of the ONF can achieve a remarkable $\eta$-value of up to 28\% \cite{le2005spontaneous,nayak2008single}. A 22\% of channeling was demonstrated by placing QDs on the surface of the ONF \cite{PhysRevLett.109.063602}. Also 7.4\% of total photons emitted from a single QD coupled to the ONF has been demonstrated \cite{fujiwara2011highly}. Single molecules have been efficiently coupled to the ONF with the maximum $\eta$-value upto 30\% \cite{skoff2018optical}. Laser cooled atoms interfaced with ONFs have also been demonstrated \cite{vetsch2010optical}. Using a flat silica ONFT, it has been numerically demonstrated that up to 38\% of photons emitted from a single quantum emitter can be directly channeled into an SMF \cite{chonan2014efficient}. Though one-sided channeling, the ONFT offers higher $\eta$-value compared to two-sided channeling in ONFs.

Cavities and hybrid systems are introduced on the ONF and the ONFT  for enhanced $\eta$-value. It has been numerically demonstrated that a maximum $\eta$-value of $\sim$75\% when an NV center is placed on a hybrid system of the ONF and a diamond nanowire \cite{yonezu2017efficient}. Single nanodiamonds containing NV centers deposited on the ONF enhances the coupling directly to an SMF \cite{schroder2011ultrabright}. The coupling of SiV centers in nanodiamonds with the ONF has been demonstrated \cite{yalla2022integration}. It has been quantitatively demonstrated that a charged exciton for a single QD deposited on the ONF and cooled down to cryogenic temperatures acts as an excellent quantum emitter \cite{shafi2018hybrid,shafi2020efficient}. Emission properties of single QD deposited on the ONF is enhanced in the presence of a single gold nanorod, leading to efficient coupling of single photons into guided modes of the ONF \cite{shafi2023bright}. QDs deposited near gold nanorods on the surface of the ONF creates an enhanced single photon source \cite{sugawara2022plasmon}. Regarding ONFTs, the enhancement in the $\eta$-value in the presence of a gold nanoparticle near the ONFT has been predicted using numerical simulations \cite{das2023efficient}. The maximum evanescent coupling of 30\% has been coupled to a tapered fiber tip from another excitation taper \cite{fielding2002tapered}. Improved antibunching using single photon sources based on statistical ensembles at the tip of the ONFT is predicted numerically and theoretically \cite{suarez2019photon}. Excitation of surface plasmons on gold-coated ONFTs and coupling them into guided mode of optical fibers have been shown experimentally \cite{auwarter2013coupling}.

Although channeling of single photons into guided modes of the ONF has been widely studied, no experimental study in this regard utilizing the ONFT has been reported. Also the fluorescence photon counts and emission spectrum measurement of the single photon emitter on the ONFT using free space and guided mode excitatioins have not been reported yet. The systematic $\eta$-value dependence study is not extended to multi-mode regime. Also, the systematic study to demonstrate the variation in the $\eta$-value according to the single quantum emitter position on the facet of the ONFT has not been reported yet.

Regarding the photon emitter, we use QDs in our experiments. QDs are semiconductor particles of a few nanometers in size \cite{jacak2013quantum,michler2000quantum}. Upon absorbing light, electrons transition from the valence band to the conduction band, generates an electron-hole pair known as an exciton. Upon recombination of the electron and hole, energy is emitted in the form of a photon. Since the QD particle size falls under the Bohr radius, the energy required to create an exciton increases, leading to the effect quantum confinement. QDs display distinctive optical characteristics due to alterations in their band gap energy induced by quantum confinement effects \cite{maxwell2020quantum}. It has been experimentally shown that CdSe/ZnS QDs exhibits a perfect antibunching at room temperature under continuous or pulsed excitation \cite{brokmann2004colloidal}.

In this paper, we demonstrate the channeling of fluorescence photons from QDs into guided modes of the ONFT. We deposit QDs on the ONFT using micro/nano fluidic technology. We measure the photon-counting and emission spectrum of fluorescence photons channeled into guided modes of the ONFT using a single photon counting module (SPCM) and a spectrometer (SP), respectively. The emission spectrum peak at $\sim$614 $nm$, with the full width at half maximum (FWHM) $\sim$22 $nm$, confirms the deposition of QDs on the ONFT. The ONFT is fabricated using a chemical etching technique employing hydrofluoric (HF) acid. We determine the $\eta$-value for the ONFT and a single dipole source (SDS) system by performing numerical simulations. We found a maximum $\eta$-value of 44\%, at the fiber size parameter ($k_{0}a$) of 7.16, corresponding to the ONFT radius of 0.71 $\mu m$. The SDS orientation and wavelength are set at radial and 620 $nm$, respectively. The SDS is placed at the center of the facet of the ONFT. Additionally, we found that sweeping the SDS position in transverse directions on the facet of the ONFT does not significantly affect the $\eta$-value. 

\section{Simulation procedure}

\begin{figure}[!h]
\centering
\includegraphics[width=\linewidth]{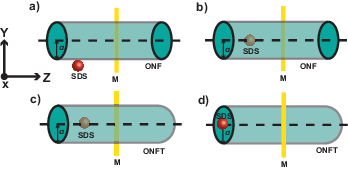}
\caption{(a) and (b) show conceptual schematics for the SDS on the surface and inside the ONF, respectively. (c) and (d) show conceptual schematics for the SDS inside and on the facet of the ONFT, respectively. ONF, ONFT, SDS, M, and {\it a} denote optical nanofiber, optical nanofiber tip, single dipole source, monitor, and radius of the ONF/ONFT, respectively.}
\label{fig1}
\end{figure}

The conceptual schematics are shown in Figs. \ref{fig1}. Numerical simulations are performed to determine the $\eta$-value for the ONF/ONFT and the SDS system using a Finite Difference Time Domain (FDTD) (FDTD package, Ansys) method \cite{gedney2022introduction,garcia2005new}. We investigate various possibilities of positioning the SDS on the ONF/ONFT and the $\eta$-value variation for each case. The system is enclosed in a three-dimensional simulation area of $6$$\times$6$\times$$25$ $\mu m^3$. The refractive index of a cylinder is chosen to be that of silica. The refractive index of the surrounding medium is set to be 1. For simulations with the ONF, the length of the silica cylinder is set at 30 $\mu m$ along the z-axis so that it crosses the perfectly matched layer (PML) boundaries and is considered infinitely long. In the case of the ONFT, the length is set at 25 $\mu m$ such that one end of the silica cylinder does not cross the PML boundary and is treated as the tip. The SDS is placed 10 $nm$ away from the surface of the ONF/ONFT to make it distinguishable for the simulation mesh and for considering the size of the QD. Note that the QD size is around 6-7 $nm$. The mesh type is set to be auto non-uniform so that the near-field SDS pattern is taken into account and boundary conditions are satisfied. The SDS wavelength is set at 620 $nm$, corresponding to the wavelength of QDs used in the present experiment. The orientation of the SDS is set along radial/azimuthal/axial direction of the ONF/ONFT. We place a 2D power monitor (M) 15 $\mu m$ away from the SDS to determine transmission ($T$) and Purcell factor ($PF$). Then, we determine $\eta$=$\frac{T}{PF}$ \cite{rao2007single,le2005spontaneous,PhysRevLett.109.063602}. The simulations are performed by varying the ONF/ONFT radius (\textit{a}) from 0.062 $\mu m$ to 1.24 $\mu m$, which corresponds to the $k_{0}a$-value from 0.62  to 12.56 ($a/\lambda$ varies from $0.1$ to $2$). For reference, we perform simulations for the following four cases

\begin{itemize}
	\item Case (i): The SDS is placed on the surface of the ONF
	\item Case (ii): The SDS is placed inside the ONF
	\item Case (iii): The SDS is placed inside the ONFT
	\item Case (iv): The SDS is placed on the facet of the ONFT
\end{itemize}

Schematics for the four cases are shown in Figs. \ref{fig1} (a)-(d). Although channeling occurs to both sides of the ONF, we show only one sided channeling for reference. In contrast, the channeling of photons occurs only to one side of the ONFT. To check the validity of simulations, the SDS is placed on the surface of the ONF, and the position of the SDS is varied along the radial direction such that the SDS moves away from the surface of the ONF to $1$ $\mu m$ away. Considering experimental ambiguity in placing QDs at the center of the facet of the ONFT, we perform simulations to determine the variation in the $\eta$-value when the SDS position is swept along the axial ($z$-axis) and transverse ($x$ and $y$-axes) directions on the facet of the ONFT. 
\section{Experimental procedure}
\subsection{Fabrication of the ONFT}
\label{sec:Fab}
The ONFT is fabricated using a two-step chemical etching technique employing HF acid (Avra, ASH2565) \cite{maruyama2006fabrication,chyad2015fabrication,kbashi2012fabrication}. We use 30\% and 24\% HF acid for the first and second steps, respectively. An etching time of 60-100 minutes for the first and 30-90 minutes for the second steps are followed. We use a silica SMF (1550B-HP, Coherent) of length 50 $cm$ for fabricating the ONFT. The etching time is controlled to produce the desired ONFT diameter. The fabricated ONFT is characterized optically and morphologically using laser and field emission scanning electron microscopy (FESEM) (Sigma 360 VP, Zeiss), respectively. Also, a quality check of the ONFT can be asserted by optical characterization. If serious scattering is observed, we don't use the sample for further experiments. Details of the fabrication procedure and characterization are reported elsewhere. The fabricated ONFT is affixed to a metal holder (MH) using a UV-curable glue for further experiments.
\subsection{Deposition of QDs on the ONFT}
\label{sec:Dep}
 \begin{figure}[!h]
 	\centering
 	\includegraphics[width=\linewidth]{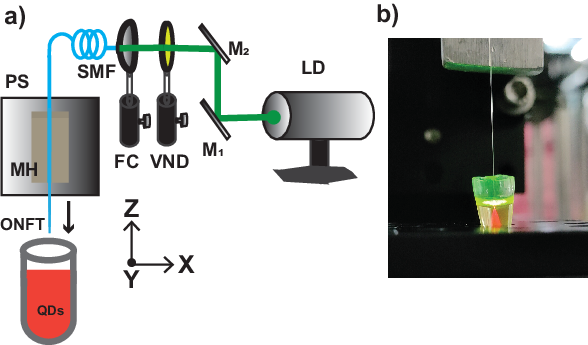}
 	\caption{(a) A conceptual diagram for the deposition of QDs on the ONFT. LD, M, VND, FC, SMF, MH, PS, ONFT, and QD denote laser diode, mirror, variable neutral density filter, fiber coupler, single mode fiber, metal holder, precision stage, optical nanofiber tip, and quantum dot, respectively. (b) The experimental picture of the scattering observed in the QD solution during deposition.}
 	\label{fig2}
 \end{figure}
As QDs are proven to be an excellent single photon emitter \cite{senellart2017high,arakawa2020progress}, we use CdSe/ZnS core-shell type colloidal QDs (900219, Sigma Aldrich) for our experiments. The original QD solution has a concentration of 5 $mg/ml$ in toluene. The original QD solution is diluted to 5\%. 5 $\mu l$ of the original QD solution is added to 95 $\mu l$ of toluene to produce 100 $\mu l$ of the diluted QD solution. 
QDs are deposited on the ONFT using micro/nano-fluidic technology.  The experimental schematic of the deposition procedure is shown in Fig. \ref{fig2} (a). The fabricated ONFT is connected to a laser diode (LD) (DL-G-5, Holmarc) at a wavelength of 532 $nm$ through two mirrors (M$_{1}$ and M$_{2}$), a variable neutral density filter (HO-VND-N50, Holmarc), and a fiber coupler (FC) (PAF2-7B, Thorlabs). For precise movement, the MH is mounted on a high-precision stage (PS) (TS 65, Holmarc). The PS is adjusted such that the ONFT is slowly dipped in the diluted QD solution. When the ONFT reaches the QD solution, one can readily observe a scattering, which serves the confirmation of the deposition, as shown in Fig. \ref{fig2} (b). Note that the scattering comes from light transmitted by the ONFT and gets enhanced when the ONFT is dipped in the QD solution. Then, the ONFT is withdrawn from the QD solution for the characterization.

\subsection{Characterization of QDs deposited ONFT}
\begin{figure}[h!]
\centering
\includegraphics[width=\linewidth]{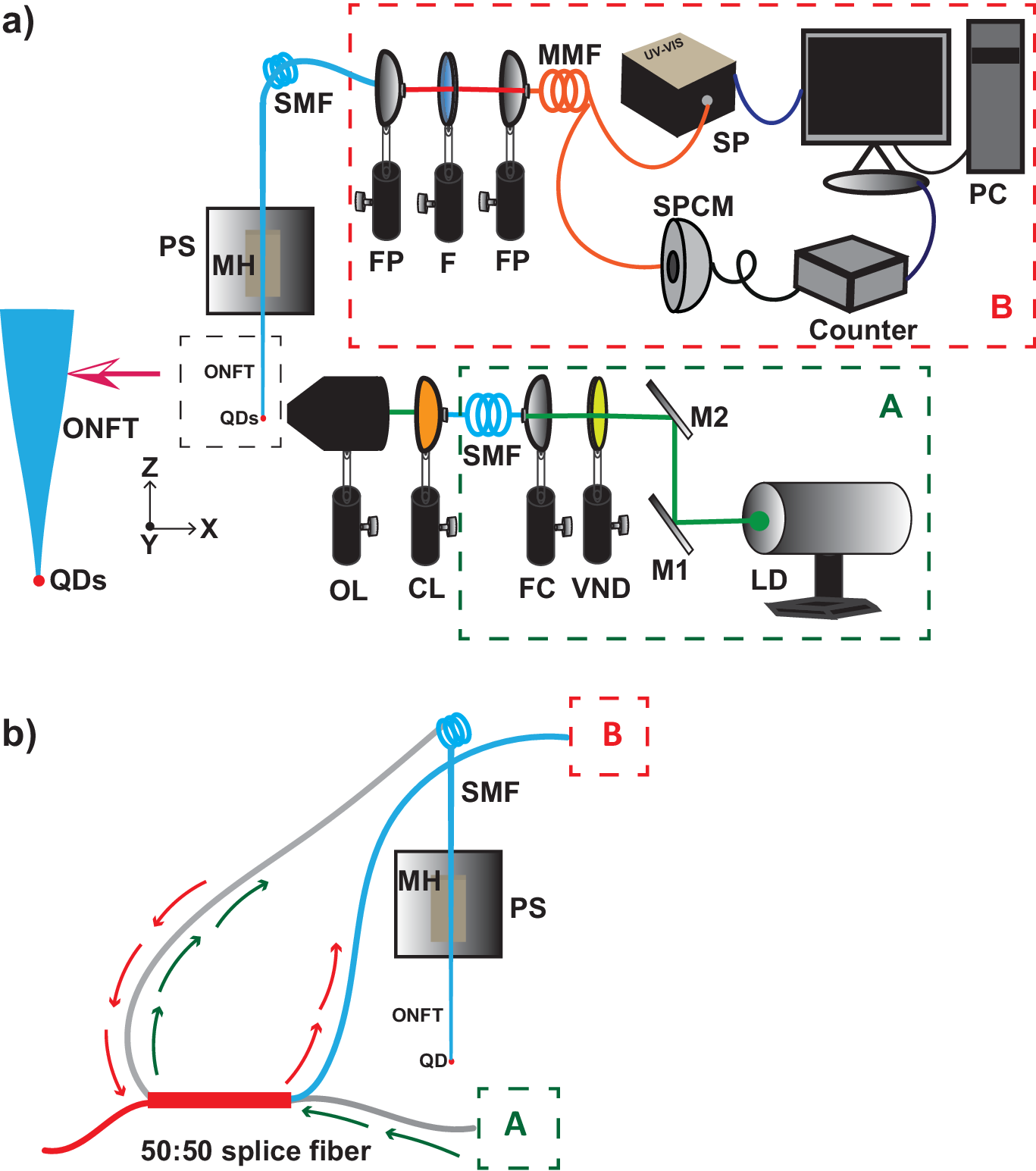}
\caption{(a) and (b) are conceptual diagrams of free space and guided mode excitations to characterize QDs deposited on the ONFT, respectively. LD, M, VND, FC, CL, OL, QD, ONFT, MH, PS, FP, F, SMF, MMF, SP, and SPCM denote laser diode, mirror, variable neutral density filter, fiber coupler, collimator, objective lens, quantum dot, optical nanofiber tip, metal holder, precision stage, fiber port, long pass filter, single-mode fiber, multi-mode fiber, spectrometer, and single photon counting module, respectively.}
	\label{fig3}
\end{figure}

We characterize QDs deposited on the ONFT in two excitation schemes. One is a free space excitation and the other is a guided mode excitation. The experimental sketch for free space excitation is shown in Fig. \ref{fig3} (a). The characterization is performed using the SP (STS-VIS, Ocean Optics) and the SPCM (SPCM-AQRH-14-FC, Excelitas). The laser light is directed to the FC as explained in section \ref{sec:Dep} and then to a collimator using an SMF. The laser light is focused on the ONFT using a microscope objective lens (OL) (Plan C N, 40X, Olympus) for exciting QDs. The PS is used to align the ONFT to the focal point of the OL. The end of the ONFT is connected to a fiber port (FP) and then passes through a long pass filter (FEL 600, Thorlabs) of cut-on wavelength 600 $nm$, and to another FP. A multi-mode fiber (MMF) is used to connect the FP to the SPCM for more signal collection. The fluorescence photons channeled into guided modes of the ONFT are measured using the SPCM connected to a photon counter (C8855-01, Hamamatsu). The emission spectrum of fluorescence photons is measured using the SP. In free space excitation, a laser power of 0.02 $\mu W$ at the SMF in Fig. \ref{fig3} (a) is focused on the ONFT using the OL. 

The experimental sketch for guided mode excitation is shown in Fig. \ref{fig3} (b). The laser light is guided through the ONFT using a 50:50 splice fiber (SF) (TW670R5A2, Thorlabs). QDs are excited through guided modes of the ONFT. The emitted fluorescence is detected through the same SF. Note that the guided mode excitation is robust against alignment compared to free space excitation. The LD is connected to one end of the SF same as for the free space excitation, mentioned as `A' in Fig. \ref{fig3} (a). The other end of the SF is connected to the SPCM/SP same as for the free space excitation, mentioned as `B' in Fig. \ref{fig3} (a). In guided mode excitation, a laser power of 0.008 $\mu W$ at the SMF in Fig. \ref{fig3} (b) is used. Here, the laser light is focused on QDs through the ONFT of diameter 312 $nm$.

\section{Simulation results} 
\begin{figure*}[t]
\centering
\includegraphics[width=0.8\linewidth]{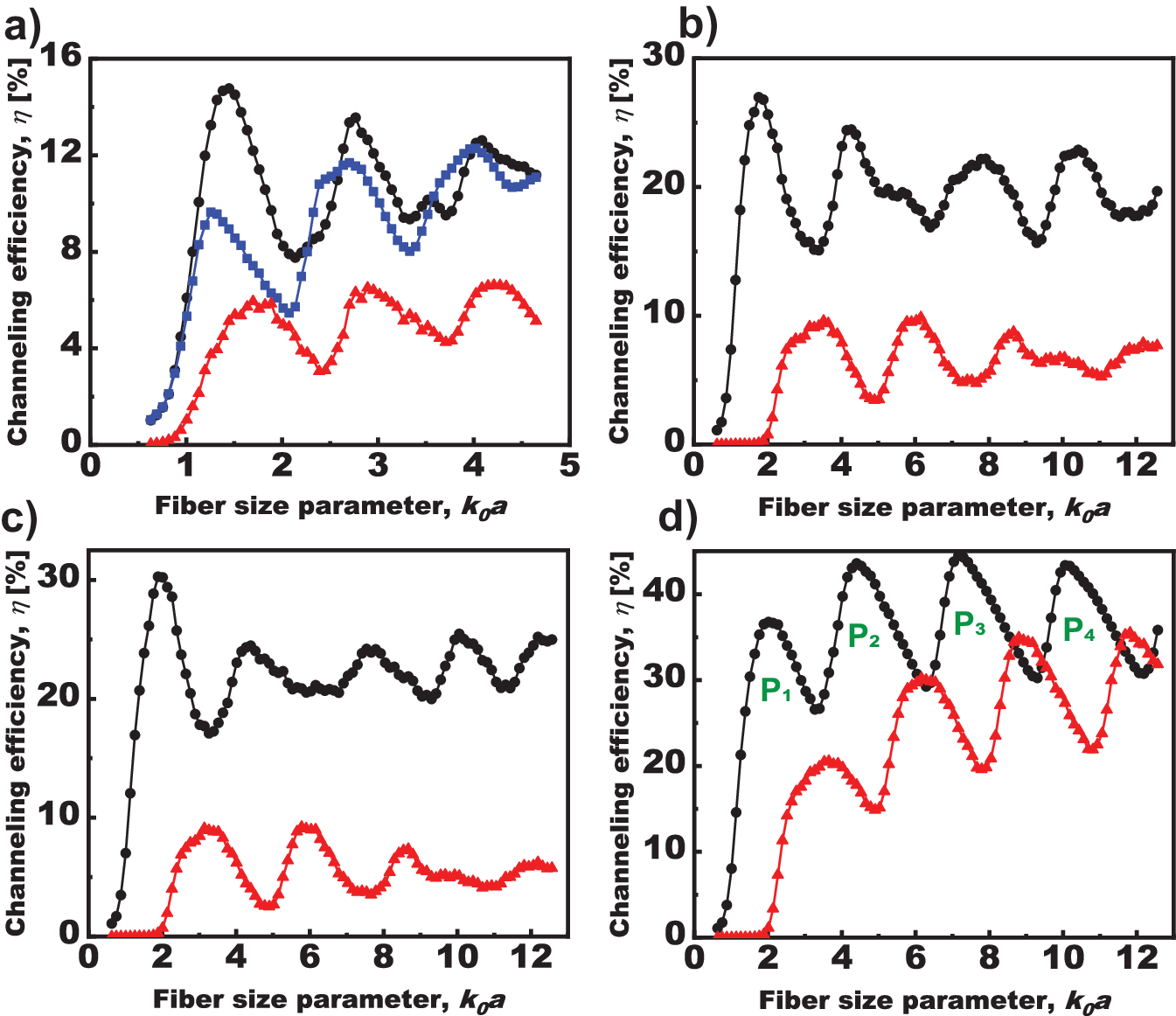}
\caption{Dependence of channeling efficiency ($\eta$) as a function of fiber size parameter ($k_{0}a$). (a) and (b) are for single dipole sources (SDS) on the surface and inside the optical nanofiber, respectively. (c) and (d) are for the SDS inside and on the facet of the optical nanofiber tip, respectively. Black circles, blue squares, and red triangles denote radial, azimuthal, and axial orientations of the SDS, respectively.}
\label{fig4}
\end{figure*}
The simulation predicted results for cases (i)-(iv) are summarised in Figs. \ref{fig4} (a)-(d). The horizontal axes denote $k_{0}a$-values and the vertical axes denote $\eta$-values, respectively. For the case (i), the $k_{0}a$-value varies from 0.62 to 4.65. For case (ii), case (iii), and case (iv), the $k_{0}a$- value varies from 0.62 to 12.56. Black circles, blue squares, and red triangles denote radial, azimuthal, and axial orientations of the SDS, respectively. One can readily observe that maximum $\eta$-value occurs for the radially orientated SDS compared to azimuthal and axial orientations for all cases. The radial and azimuthal orientations are same for Figs. \ref{fig4} (b)-(d). Note that $\eta$-values are for one side of the ONF. For the case (i) and (ii), the peak $\eta$-value of 15\% and 27\% occurred for the radial orientation at the $k_{0}a$-value of 1.44 and 1.76, which corresponds to the ONF radius of 0.14 $\mu m$ and 0.17 $\mu m$, respectively. For case (ii), one can observe that the maximum $\eta$-value is almost twice that of  case (i). 

For the case (iii), the peak $\eta$-value of 30\% occurred at the $k_{0}a$-value of 1.88, corresponding to the ONFT radius of 0.18 $\mu m$. It is obvious from the results that when the ONFT is used instead of the ONF, there is an increase in the $\eta$-value. For the case (iv), one can readily see four peaks. We denote the first, second, third, and fourth peaks as P$_{1}$, P$_{2}$, P$_{3}$, and P$_{4}$, respectively. The maximum $\eta$-value corresponds to P$_{3}$ in Fig. \ref{fig4} (d). For P$_{3}$, the maximum $\eta$-value of 44.5\% is obtained at the $k_{0}a$-value of 7.16, which corresponds to the ONFT radius of 0.71 $\mu m$. The P$_{1}$, P$_{2}$, and P$_{4}$ $\eta$-values of 37\%,  43.5\%, and 43\% occurred at $k_{0}a$-values of 2.01, 4.39, and 10.05, which corresponds to the ONFT radius of 0.2 $\mu m$, 0.43 $\mu m$, and 1 $\mu m$, respectively. 

\begin{figure*}[ht]
	\centering
	\includegraphics[width=0.8\linewidth]{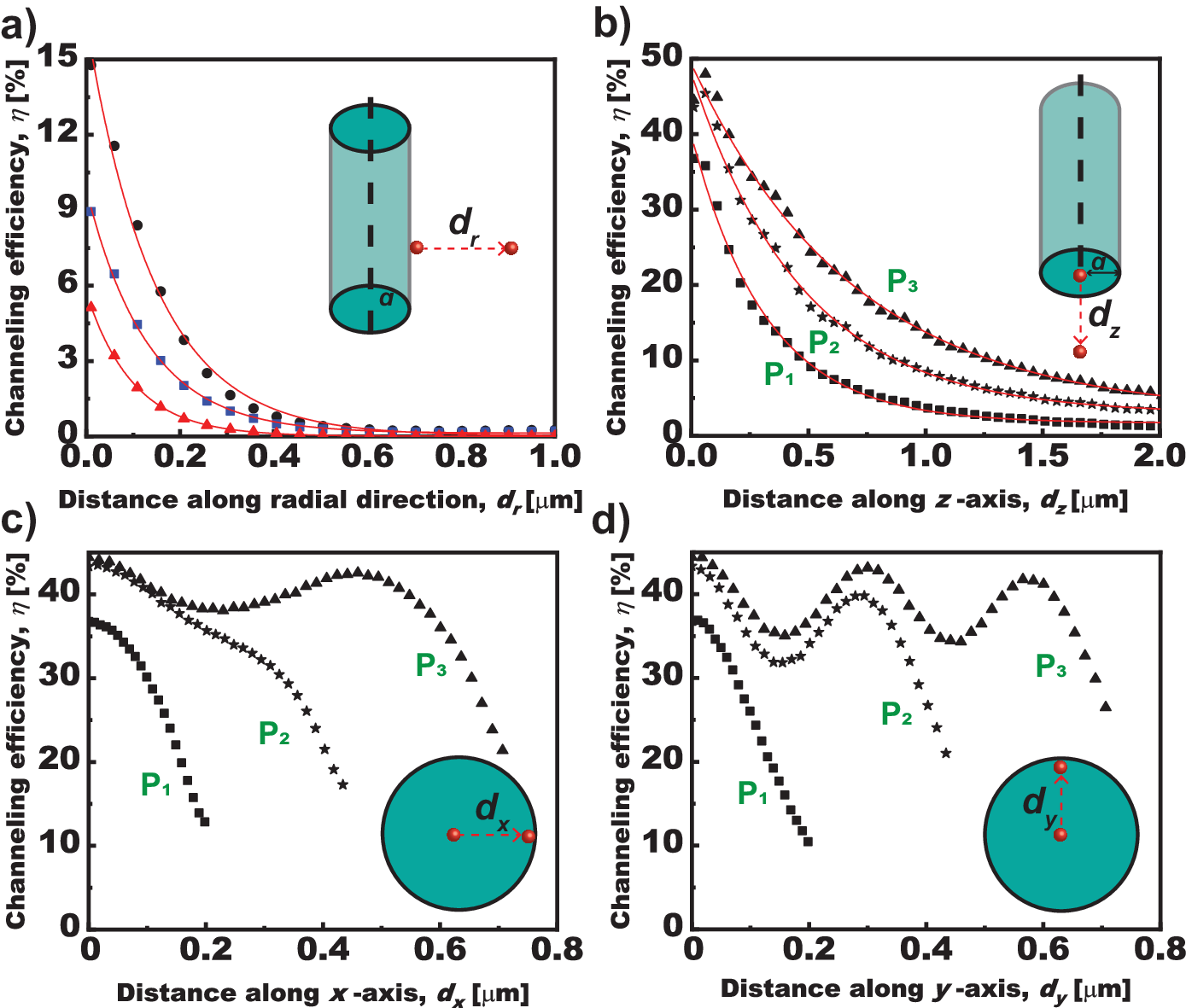}
	\caption{The dependence of channeling efficiency ($\eta$) as a function of the single dipole source (SDS) position: (a) is for the radial direction of the optical nanofiber. Black circles, blue squares, and red triangles represent radial, azimuthal, and axial orientations of the SDS. (b), (c), and (d) are for $z$, $x$, and $y$-directions on the facet of the optical nanofiber tip, respectively. Black squares, stars, and triangles represent $k_{0}a$- values of 2.01, 4.39, and 7.16, respectively. (b), (c), and (d) correspond to the radial orientation of the SDS. Insets show the position of the SDS.}
	\label{fig5}
\end{figure*}

Next, we investigate the SDS position dependence on the $\eta$-value from the surface along the radial direction of the ONF. The corresponding simulation predicted result is shown in Fig. \ref{fig5} (a). The $k_{0}a$-value of the ONF is set at 1.44. The horizontal axis denotes distance along the radial direction ($d_{r}$) and the vertical axis denotes $\eta$-value. Black circles, blue squares, and red triangles represent radial, azimuthal, and axial orientations of the SDS, respectively. Data points are fitted with a single exponential decay function, denoted by red solid line. As the SDS moves away from the ONF, the $\eta$-value decreases exponentially. 

Additionally, we sweep the SDS position along $z$, $x$, and $y$ directions from the center of the facet of the ONFT. The simulation predicted results are summarised in Figs. \ref{fig5} (b)-(d). The horizontal axes denote distance along the $z$-direction ($d_{z}$)/$x$-direction ($d_{x}$)/$y$-direction ($d_{y}$) and the vertical axes denote $\eta$-values. Black squares, black stars, and black triangles represent $k_{0}a$- values 2.01 (P$_{1}$), 4.39 (P$_{2}$), and 7.16 (P$_{3}$), respectively. All the results correspond to the radial orientation. The simulation predicted results for sweeping the SDS position along the $z$-direction of the ONFT are shown in Fig. \ref{fig5} (b). Data points are fitted with a single exponential decay function, denoted by the red solid line. As the SDS goes away from the ONFT, the $\eta$-value decreases. The simulation result corresponding to sweeping the SDS position along the $x$-axis on the facet of the ONFT is shown in Fig. \ref{fig5} (c). One can readily observe that the $\eta$-value does not vary significantly for the $k_{0}a$-value of 7.16. The maximum/minimum $\eta$-value is 44.5\%/21\%.  From the center of the facet to 0.5 $\mu m$ away along the $x$-direction, the $\eta$-value of 40\% is being maintained. The $\eta$-value variation for $k_{0}a$-values 2.01 and 4.39 is significant. For the $k_{0}a$-values of 2.01 and 4.39, the maximum/minimum $\eta$-values are 37\%/13\% and 43.5\%/17\%, respectively. The simulation result corresponding to sweeping the SDS position along the $y$-axis on the facet of the ONFT is shown in Fig. \ref{fig5} (d). One can readily observe that the $\eta$-value does not vary significantly for the $k_{0}a$-value of 7.16. The maximum/minimum $\eta$-value is 44.5\%/26\%.

\section{Experimental results}
\begin{figure}[!h]
	\centering
	\includegraphics[width=0.9\linewidth]{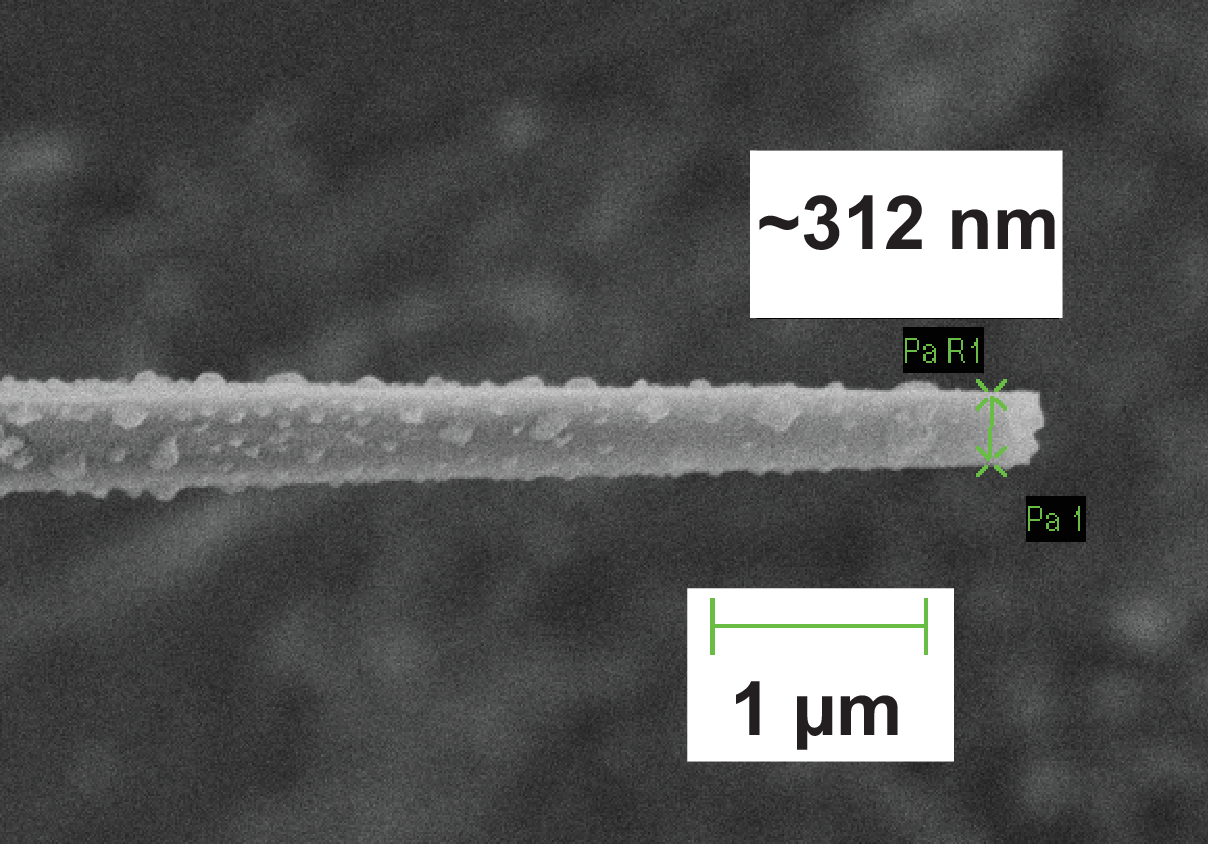}
	\caption{A typical field emission scanning electron microscopy image of an optical nanofiber tip diameter of $\sim$312 $nm$.}
	\label{fig6}
\end{figure}

\begin{figure*}[ht]
	\centering
	\includegraphics[width=0.8\linewidth]{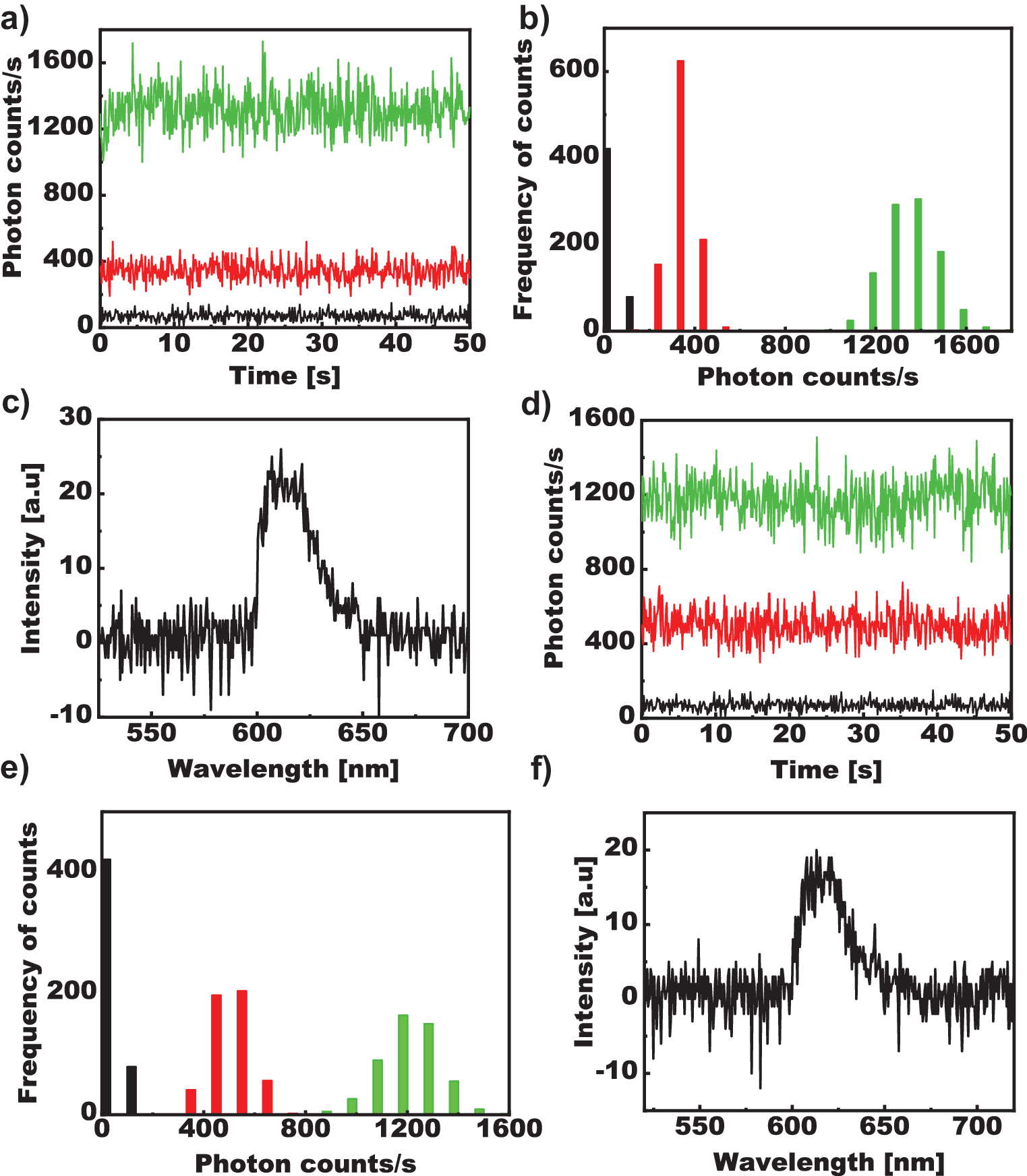}
	\caption{(a) and (d) are measured photon counts as a function of time using free space excitation and guided mode excitation, respectively. (b) and (e) are histograms corresponding to (a) and (d), respectively. (c) and (f) are background-corrected fluorescence spectra measured using free space excitation and guided mode excitation, respectively. Black, red, and green traces/ bars denote the dark count rate of the single photon counting module, background, and fluorescence photon counts from quantum dots into guided modes of the optical nanofiber tip, respectively.}
	\label{fig7}
\end{figure*}

A typical FESEM image of a fabricated ONFT is shown in  Fig. \ref{fig6}. The diameter of the ONFT is found to be $\sim$0.31 $\mu m$. The measured optical transmission is $\sim$50\%. We performed the photon counting experiment and spectrum measurement for 15 samples. Out of these, only typical spectrum and photon counts are shown here. A typical observed photon counting as a function of time using free space excitation is shown in Fig. \ref{fig7} (a). The horizontal axis denotes time in seconds, and the vertical axis denotes counts (c/s). Black, red, and green traces denote the dark count rate of the SPCM, background, and fluorescence photon counts from QDs into the ONFT, respectively. The corresponding histogram for the three is shown in Fig. \ref{fig7} (b). An average of 70 ($\pm$25) c/s and 350 ($\pm$60) c/s are observed as the dark count rate of the SPCM and background, respectively. An average of 1312 ($\pm$120) c/s is observed as the fluorescence photon counts when QDs on the ONFT are focused. The background corrected fluorescence spectrum observed using free space excitation is shown in Fig. \ref{fig7} (c). It is fitted with a Lorentzian peak function. It exhibits a peak wavelength at 614 $nm$ with FWHM 22 ($\pm$1) $nm$.

A typical observed photon counting as a function of time using guided mode excitation is shown in Fig. \ref{fig7} (d). The horizontal axis denotes time in seconds and the vertical axis denotes counts (c/s). Black, red, and green traces denote the dark count rate of the SPCM, background, and fluorescence photon counts from QDs into the ONFT, respectively. The corresponding histogram for the three is shown in Fig. \ref{fig7} (e). An average of 70 ($\pm$25) c/s, 498 ($\pm$75) c/s, and 1171 ($\pm$110) are observed as the dark count rate of the SPCM, background, and fluorescence photon counts from QDs on the ONFT, respectively. The background-corrected fluorescence spectrum observed using guided mode excitation is shown in Fig. \ref{fig7} (f). It is fitted with a Lorentzian peak function. It exhibits a peak wavelength at 617 $nm$ with FWHM 22 ($\pm$1) $nm$.
\section{Discussions}
As seen in Fig. \ref{fig4} (a), the result indicates that the maximum $\eta$-value is 15\%. The predicted results are in good agreement with previously reported results \cite{PhysRevLett.109.063602,le2005spontaneous}. When the SDS is placed inside the ONF, there is a significant increase in the $\eta$-value to 27\% as shown in Fig.\ref{fig4} (b). It is consistent with previously reported results \cite{takezawa2023room}. This clearly implies that when the SDS is positioned inside the ONF instead of on the surface, the $\eta$-value is almost doubled. Although the $\eta$-value is higher than that of the case (i), it is less than 30\% when the ONF is used. The $\eta$-value crossed 30\% for the case (iii) as shown in Fig. \ref{fig4} (c). This simulation result implies that the $\eta$-value for the ONFT is higher than the ONF. The present $\eta$-value of 44\% is achieved when the SDS is positioned on the center of the facet of the ONFT as shown in Fig. \ref{fig4} (d). This signifies that, in this configuration, a substantial number of photons emitted from the SDS effectively channels into guided modes of the ONFT. Due to the interference between the radiation from the SDS and that from the mirror image of the SDS on the facet of the ONFT, the $\eta$-value is enhanced compared to the cases where the SDS is on the surface or inside the ONF/ONFT \cite{chonan2014efficient}.

As seen in Fig. \ref{fig4} (d),  the single mode condition for the wavelength 620 $nm$ is satisfied when $k_{0}a$$<$ $2.28$. The first peak $\eta$-value ($k_{0}a$= $2.01$) corresponds to the single-mode regime, and the successive peaks correspond to the multi-mode regime. For the first peak, the contribution is only from the fundamental mode ($HE_{11}$) and for the other three peaks contribution is from the fundamental mode and higher order modes, leading to an increase in the $\eta$-value. The maximum $\eta$-value of 44.5\% occurred at the $k_{0}a$-value of 7.16, which corresponds to the ONFT radius of 0.71 $\mu m$. This result indicates that the SMF is not necessarily tapered to sub-wavelength scales to realize the maximum $\eta$-value. This implies that simulated results can be feasible for experiments. The present guided mode excitation experiments are performed in the single-mode regime since the ONFT diameter of 0.31 $\mu m$ corresponds to the $k_{0}a$-value of 1.57. Experiments in the multi-mode regime can be done as it is easy to fabricate thicker ONFTs.

As seen in Fig. \ref{fig5} (a), as the SDS moves away from the ONF, the $\eta$-value decreases exponentially as expected. As the SDS moves away from the ONF/ONFT, evanescent field gets decayed and hence $\eta$-value follows exponential decay. The predicted results are in good agreement with previously reported results \cite{le2005spontaneous,le2004field}. As seen in Fig. \ref{fig5} (b),  if another single photon source is used instead of QDs, its size variation can be accounted. One can observe that the $\eta$-value pattern for the P$_{1}$ in Figs. 5 (c) and (d) are almost the same. This is due to the operational condition in the single-mode regime.  In contrast, for the other two peaks, P$_{2}$ and P$_{3}$, which are in the multimode regime, there is an oscillatory behavior. This may be due to the asymmetry in the multimode field distribution along x and y-directions \cite{verhart2014single}. Note that, in both cases, the SDS is oriented along the radial direction. As seen in Fig. \ref{fig5} (c), from the center of the facet to 0.5 $\mu m$ away along $x$-direction, the $\eta$-value of 40\% is being maintained for the $k_{0}a$-value of 7.16. This implies that the SDS is not necessarily to be placed at the center of the facet of the ONFT. From the experimental point of view, it gives the freedom to deposit QDs anywhere along the $x$-direction on the facet. But the SDS position along the $y$-direction has to be determined carefully as seen in Fig. \ref{fig5} (d).

Though irregularities exist on the ONFT shown in Fig. \ref{fig6}, its size is much smaller than the wavelength detected, 620 $nm$. Therefore it does not seriously affect the light confinement. As seen in Fig. \ref{fig7} (a), the measured dark count rate of the SPCM is 70 ($\pm$25) c/s, which is consistent with the value quoted by the company. As seen as red traces, 350 ($\pm$60) c/s is observed as the background photon counts, which includes the dark counts of the SPCM. As seen as green traces, an average of 1312 ($\pm$120) c/s is observed as the fluorescence photon counts. One can understand that only a total of $962$ ($\pm$134) c/s is the fluorescence photon counts from QDs that are channeled into guided modes of the ONFT. In a similar fashion, as seen in Fig. \ref{fig7} (d), an average of 673 ($\pm$133) c/s is observed as the fluorescence photon counts. Note that the ONFT used for free space and guided mode excitations are different. Also considering all losses through the guided mode excitation, the observed counts is much less than the actual fluorescence photons counts. The guided mode excitation undergoes high coupling loss comapared to the free space excitation. Note that keeping the ONFT aligned with the focal point of the OL throughout the experiment is challenging. Hence, guided mode excitation is preferred over free space excitation to avoid mechanical instability in the system. As the simulation predicted, a significant amount of photons are channeled into the ONFT. The peak wavelength in the emission spectrum of fluorescence photons from QDs channeled into guided modes of the ONFT confirms the deposition of QDs on the ONFT. The peak matches with the company-quoted emission wavelength 620 ($\pm$10) $nm$. Hence, we demonstrated the channeling of fluorescence photons from QDs into guided modes of the ONFT. 

One can quantitatively measure the $\eta$-value by observing the photon counting rate for guided modes and radiation modes simulatenously. The $\eta$-value of 22\% is already experimentally demonstrated using the ONF and a single QD system \cite{PhysRevLett.109.063602}. Therefore, it should be possible to quantitatively measure the $\eta$-value with the ONFT. As described in \ref{sec:Dep}, we dip the ONFT in the QD solution, leading to ambiguity in the position of deposition of QDs. We do further experiments to investigate the exact location of QDs deposited. The ongoing experiments to realize the exact location of QDs deposited on the ONFT and the number of QDs deposited will be discussed elsewhere. The deposition of a single QD on the surface of the ONF has been realized and reproduced theoretical predictions \cite{PhysRevLett.109.063602}. Therefore, we believe that a single QD on the ONFT is also possible to realize and reproduce the present simulation results. The ongoing experiments regarding this will be reported elsewhere. Although we use QDs in our experiments, the study can be extended for atoms and defects in nanodiamonds.
\section{Summary}
In summary, we demonstrated the channeling of fluorescence photons from quantum dots (QDs) into guided modes of an optical nanofiber tip (ONFT). We deposited QDs on the ONFT using micro/nano fluidic technology. We measured the photon-counting and emission spectrum of fluorescence photons channeled into guided modes of the ONFT using two excitation schemes- free space and guided mode.  The measured emission spectrum confirms the deposition of QDs on the ONFT. We determined channeling efficiency ($\eta$) for the ONFT and a single dipole source (SDS) system by performing numerical simulations. The maximum $\eta$-value of 44\% occurred at the fiber size parameter of 7.16, corresponding to the ONFT radius of 0.71 $\mu m$ for the emission wavelength at 620 $nm$. Additionally, we investigated the SDS position dependence in transverse directions on the facet of the ONFT in view of keeping experimental ambiguities. The automatic fiber coupling scheme is a unique aspect of the present scheme. This cavity-free and operating at room temperature method is a unique development to realize a quantum network node. The present work will have significant applications in modern physics, chemistry, biology and industry. The quantum functionalized optical fibers can be readily used to create the quantum internet. The present fiber inline platform may open new avenues in quantum technologies.

\section*{Acknowledgments}
RM acknowledges the University Grants Commission (UGC) for the financial support (Ref. No.:1412/CSIR-UGC NET June 2019). RRY acknowledges financial support from the Science and Engineering Research Board (SERB) for the Core Research Grant (CRG) (File No. CRG/2021/009185), Institute of Eminence (IoE) grant at the University of Hyderabad, Ministry of Education (MoE) (File No. RC2-21-019), and the Scheme for Transformational and Advanced Research in Sciences (STARS) grant from Ministry of Human Resource Development (MHRD) (File No. STARS/APR2019/PS/271/FS).
\section*{References}
\bibliographystyle{iopart-num.bst}
\bibliography{iopart-num}
\end{document}